\documentclass{aa}
\usepackage{graphicx}
\usepackage{natbib}

\usepackage{txfonts}
\begin{document}
\title{Orbital motion of the young brown dwarf 
companion TWA 5 B \thanks{Based on observations 
collected at the European Southern Observatory, Chile, in 
runs 79.C-0103(A) and 81.C-0393(A) as well as on data
obtained from the public ESO science archive.}}

\author{Ralph Neuh\"auser\inst{1} \and
Tobias O.B. Schmidt\inst{1} \and
Valeri V. Hambaryan\inst{1} \and
Nikolaus Vogt\inst{2,3} 
}

\institute{
Astrophysikalisches Institut, Universit\"at Jena, Schillerg\"asschen 2-3, 07745 Jena, Germany
\and
Departamento de F\'isica y Astronom\'ia, Universidad de Valpara\'iso,
Avenida Gran Breta\~na 1111, Valpara\'iso, Chile
\and
Instituto de Astronomia, Universidad Catolica del Norte, Avda. Angamos 0610,
Antofagasta, Chile
}

\offprints{Ralph Neuh\"auser, \email{rne@astro.uni-jena.de}}

   \date{Received 21 Dec 2009; accepted 4 March 2010 }

  \abstract
{It is difficult to determine masses and test formation models for brown dwarfs,
because they are always above the main sequence, so that there is a degeneracy between
mass and age. However, for brown dwarf companions to normal stars,
such determinations may be possible, because one can know the distance and 
age of the primary star. 
As a result, brown dwarf companions are well-suited to testing formation 
theories and atmosphere models.
}
{With more adaptive optics images available, we aim at detecting 
orbital motion for the first time in the system TWA 5 A+B.
} 
{We measured separation and position angle between TWA 5 A and B in each high-resolution
image available and followed their change in time, because B should orbit around A.
The astrometric measurement precision is about one milli arc sec.
}
{With ten year difference in epoch, we can clearly detect orbital motion of B
around A, a decrease in separation by $\sim 0.0054^{\prime \prime}$ per year
and a decrease in position angle by $\sim 0.26^{\circ}$ per year.
}
{TWA 5 B is a brown dwarf with $\sim 25$ Jupiter masses (Neuh\"auser et al. 2000), 
but having large error bars (4 to 145 Jupiter masses, Neuh\"auser et al. 2009).
Given its large projected separation from the primary star, $\sim 86$ AU, 
and its young age ($\sim 10$ Myrs), it has probably 
formed star-like, and would then be a brown dwarf companion.
Given the relatively large changes in separation and position angle 
between TWA 5 A and B, we can conclude that they orbit around each other 
on an eccentric orbit. Some evidence is found for a curvature in the orbital 
motion of B around A - most consistent with an elliptic (e=0.45) orbit.
Residuals around the best-fit ellipse are detected and show a small-amplitude 
($\sim 18$ mas) periodic sinusoid with $\sim 5.7$ yr period, 
i.e., fully consistent with the orbit of the inner close pair TWA 5 Aa+b.
Measuring these residuals caused by the photocenter wobble - even in unresolved 
images - can yield the total mass of the inner pair, 
so can test theoretical pre-main sequence models.
}

\titlerunning{Orbital motion of TWA 5 B}


\maketitle

\section{Introduction: The brown dwarf TWA 5 B}

The star TWA 5 is one of the five original members
of the TW Hya association (TWA), 
a group of 5 to 12 Myr young stars (Kastner et al. 1997),
where no gas clouds are left from the star formation process (Tachihara et al. 2009);
see Torres et al. (2008) for a recent review on TWA. TWA 5 is an M1.5 weak-line 
T Tauri star (Webb et al. 1999)
with variable H$\alpha$ emission, hence still ongoing accretion (Mohanty et al. 2003). 
The central star itself is either a very close ($\le 66$ milli arc sec or mas) binary 
(Konopacky et al. 2007, henceforth K07) or even triple (Torres et al. 2003).
The close inner pair TWA 5 Aa+b has a total dynamical mass of $0.71 \pm 0.14$~M$_{\odot}$
(assuming 44 pc as distance) and an orbital period of $5.94 \pm 0.09$ years (K07).
The wide companion TWA 5 B was originally discovered by
Webb et al. (1999) and Lowrance et al. (1999) and confirmed as
co-moving with TWA 5 A by Neuh\"auser et al. (2000).

The spectral type of TWA 5 B is M8-9 (Webb et al. 1999, Lowrance et al. 1999,
Neuh\"auser et al. 2000, Mohanty et al. 2003).
The mass of the companion is between 15 and 40 
Jupiter masses just from temperature, luminosity, and
theoretical hot-start model tracks (Neuh\"auser et al. 2000).
The mass lies anywhere between 4 and 145 Jupiter masses, if calculated 
from temperature ($2800 \pm 100$ K), 
luminosity ($\log(L_{bol}/L_{\odot}) = -2.62 \pm 0.30$ at $44 \pm 4$ pc), 
and gravity ($\log g = 4.0 \pm 0.5$),
as obtained by comparison of a 
Sinfoni K-band spectrum with Drift-Phoenix model atmospheres 
(Neuh\"auser et al. 2009).

The system TWA 5 A+B was observed by several teams
with ground-based adaptive optics (AO) and/or the Hubble Space Telescope (HST).
We obtained two more recent images, so that we
can now investigate possible orbital motion of B around A with a 10 year 
difference in epoch (first preliminary results in Schmidt et al. 2008).
We can then also try to detect the orbital motion of Ab and Aa
around each other as residuals of the much longer orbit of B around A
due to a periodic wobble of the photocenter of the close Aa+Ab pair.

\section{Astrometry with VLT/NACO}

We observed TWA 5 with the adaptive optics imager NACO 
(for NAOS CONICA for the Nasmyth Adaptive Optics System, NAOS, with
the COude NearInfrared Camera and Array, CONICA, Rousset et al. 2003) 
at the ESO VLT in 2007 and 2008 with the S13 camera, i.e.,
a $14^{\prime \prime} \times 14^{\prime \prime}$ field of view.
In July 2007, we obtained AO images in the following jitter set-up.
Each individual image had a detector integration time (DIT) of 30 sec;
the number of DITs (NDIT) co-added together immediately after exposure,
i.e. without shifting, but added up and saved in one single file, was 3,
resulting in 90 sec total exposure per fits file; and the number of
such integrations (NINT) was 14, so that we 
had $30 \times 3 \times 14$ sec = 21 min total integration time.
We always used the neutral density filter,
because of the brightness of TWA 5 and the good seeing ($0.6^{\prime \prime}$).
On 12 June 2008, we obtained four (NINT) times 174 (NDIT) times 0.3454 sec (DIT) images
under less good conditions, repeated in the next night under better conditions
(with NINT 5), i.e., 9 min total integration time.

All science and flat-field frames taken were subtracted by a dark frame,
then the science frames were divided by a normalized flat field.
A shift+add procedure was applied to subtract the background and
to add up all frames for each run. We used ESO eclipse and MIDAS.
The same procedure was performed for the astrometric standard star binary HIP 73357.
Astrometric data on separations and position angles (PA) include
Gaussian centering errors in the science targets
and the astrometric standards, as well as possible motion in the standards
(see e.g. Neuh\"auser et al. 2008 for details on typical astrometric
calibration procedure and maximum orbital motion in HIP 73357). 

\begin{figure}
\includegraphics[angle=270,width=1\hsize]{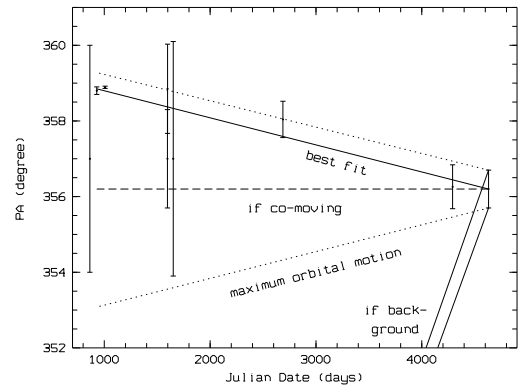}
\caption{Position angle (in degrees) versus observing epoch (JD - 2450000 in days) 
for {\em corrected} data listed in Table 1. The dotted lines (starting from the
2008 data point opening to the past) 
indicate maximum PA change due to orbital motion
for a circular pole-on orbit. The full lines with strong positive
slope in the lower right corner are for the background
hypothesis, if the bright central star (TWA 5 A) had 
according to its known proper motion, while the fainter 
northern object (now known as B) would be a non-moving object;
the data points are inconsistent with the background hypothesis 
by many $\sigma$.
All data points are fully consistent with common proper
motion, but not exactly identical proper motion (plotted as dashed line).
Instead, the data are fully consistent with orbital motion
(inclination between pole-on and face-on),
because the PA appears to decrease by $\sim 0.26^{\circ}$ per year
(plotted as a best fit, full line).}
\end{figure}

\begin{figure}
\includegraphics[angle=270,width=1\hsize]{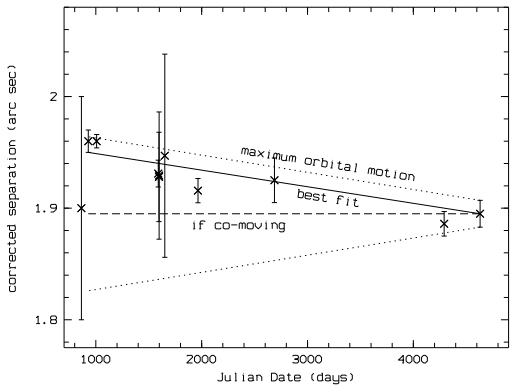}
\caption{Separation (in arc sec) versus observing epoch (JD - 2450000 in days) 
for {\em corrected} data listed in Table 1.
The dotted lines (starting from the 2008 data point opening to the
past) indicate maximum possible separation change due to orbital motion
for a circular edge-on orbit. The expectation for the background hypothesis
is not shown for clarity and because it was already rejected in the previous figure.
All data points are fully consistent with common proper motion, 
but not exactly identical proper motion (plotted as dashed line).
Instead, the data are fully consistent with orbital motion
(inclination between pole-on and edge-on),
because the separation decreases by $\sim 5.4$ mas per year
(plotted as a best fit, full line).}
\end{figure}

\begin{figure}
\includegraphics[angle=0,width=1\hsize]{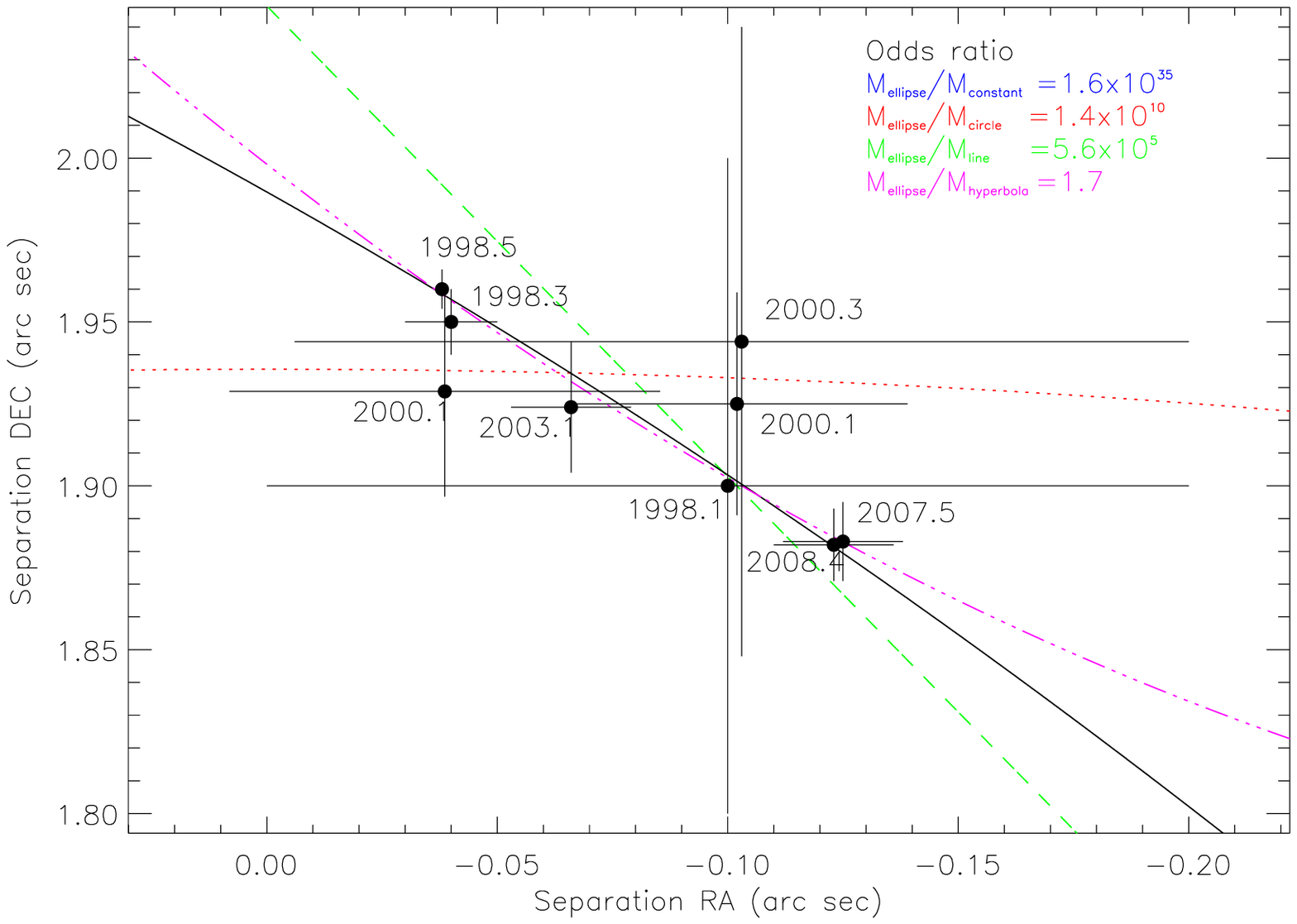}
\caption{Separations in $\delta$ versus $\alpha$ (both in arc sec) 
for {\em corrected} data listed in Table 1.
The brown dwarf B moves to the south-west (relative to A).
We tried to fit a geometric orbit for different models of motion of B wrt A:
Solid (black) line for an elliptic orbit of B around A (best fit),
dotted (red) line for a circular orbit of B around A,
dashed (green) line for constant change in separation in $\alpha$ and $\delta$
(as if B would be a background object, i.e. with negligible parallax for B,
with proper motion of B being different from A,
and taking into account parallactic motion of TWA 5 A at
a distance of $44 \pm 4$ pc; for clarity, we do not plot the yearly parallactic
wobble (for $\pm 44$ pc) and its uncertainty ($\pm 4$ pc),
but show the relevant model as averaged straight line motion),
and dash-dotted (magenta) line for hyperbolic motion of B wrt A (i.e. B being ejected);
an additional model (constant separation in $\alpha$ and $\delta$) is not plotted.
We list the odds ratios from Bayesian statistics in the upper right:
The ellipse is 2 to 3 times more likely than the constant change (background) and 
the hyperbola (ejected), respectively. While this is only a geometric fit,
we point out that the first two data points with small errors (1998 from HST)
are in the upper left (NE), the data from the middle epochs (2000) are in
the center of the plot (large errors), and the last two data points with small errors 
(2007 and 2008 from NACO) are in the lower right (SW), i.e. following
the (geometric) orbit. }
\end{figure}

The NACO S13 pixel scale for 2007 July 8 was determined to be $13.264 \pm 0.079$ 
mas/pixel with the detector orientation shifted by $0.30 \pm 0.40 ^{\circ}$;
the pixel scale for 2008 June 13 was determined to be $13.243 \pm 0.086$ mas/pixel
with the detector orientation shifted by $0.73 \pm 0.40 ^{\circ}$.
These orientation values have to be added to a value measured on a raw frame.
For the 2003 NACO data by Masciadri et al. (2005), which we reduced again,
no astrometric calibration targets were observed, so that we obtained the
(rough) calibration from the fits file headers (from the position keywords
compared to the star position in the image), namely 
$13.22 \pm 0.13$ mas/pixel (close to the nominal value)
and a detector orientation $0.30 \pm 0.37^{\circ}$.
In Table 1, we list all imaging observations used here with
separations and angles measured.

TWA 5 A is itself a close visual pair,
where Aa is slightly brighter than Ab (Macintosh et al. 2001,
Brandeker et al. 2003, K07).
The close pair Aa+b (separation $\le 66$ mas, see K07) 
is not resolved in most of the observations listed 
in Table 1 (except the two obtained with Keck in 2000.1). 
We thus have to correct the separation of B wrt A for
the photocenter motion of Aa+b. K07 have solved
the orbit of Ab around Aa, which we can use for 
the correction.\footnote{There are two typos in Table 1 in K07:
The position angle of the Aa+b pair should be $25.9 \pm 1.0 ^{\circ}$ 
for the 2000 Feb 20 data from Macintosh et al. (2001), 
who gave $25.9 \pm 0.5 ^{\circ}$; and it should be $24.2 \pm 3.0 ^{\circ}$ for
the 22 Feb 2000 data from Brandeker et al. (2003), who gave 
$24.15 \pm 2.8 ^{\circ}$; the Feb 2000 positions of Ab relative to Aa
are correctly plotted in Fig. 3 in K07; however, 
the signs of the RA values given in Fig. 3 in K07 are wrong.
Since the slightly fainter Ab is located
towards the NE in Feb 2000 (Macintosh et al. 2001, Brandeker et al. 2003)
and Dec 2005 (K07), the RA axis in Fig. 3 in
K07 should go from west (left, negative RA changes)
to east (right, positive RA changes).}
That Aa+b was not resolved in the HST/Nicmos images in 
1998.3 (aquisition without coronograph) and 1998.5 is consistent 
with Aa+b being close to their closest approach, as suggested 
by Macintosh et al. (2001) and consistent with the orbit of 
K07. The closest approach was in 1998.4 with a few mas separation.

We used the orbit in K07 to correct the
separations measured between A and B for the photocenter shift
in Aa+b, assuming that Aa and Ab are equally bright. The K-band 
magnitude difference is only $0.23 \pm 0.09$ mag (K07).
TWA 5 Ab orbits Aa with $5.94 \pm 0.09$ yr period and
a semi-major axis of $0.066 \pm 0.005 ^{\prime \prime}$ (K07).
The correction for the epochs listed in Table 1 amounts to
$\sim 28$ mas, so is significant.
The corrected values are also listed in Table 1.

\begin{table*}
\begin{tabular}{lllllllll}
\multicolumn{9}{c} Table 1. Astrometry of TWA 5 A and B \\ \hline
Epoch  & JD -    & Telesope and & Band & \multicolumn{3}{c}{separation [arc sec] (a)} & PA (b)       & Ref. \\
year   & 2450000 & instrument   &     & in RA & in Dec & total & [$^{\circ}$] & (c)  \\ \hline

1998.1 & 863.5  & IRTF Speckle  & K   & $-0.1 \pm 0.1$ & $1.9 \pm 0.1$ & $1.9    \pm 0.1    $&$ 357     \pm 3   $ & Webb99 \\
\multicolumn{4}{r}{corrected:} & $-0.1 \pm 0.1$ & $1.9 \pm 0.1$ & $1.9    \pm 0.1    $&$ 357     \pm 3   $ & \\
1998.3 & 928.8  & HST/Nicmos    & H   &$-0.04 \pm 0.01$ & $1.95 \pm 0.01$ & $1.96   \pm 0.01   $&$ 358.8   \pm 0.1 $ & Low99(f) \\
\multicolumn{4}{r}{corrected:} & $-0.04 \pm 0.01$ & $1.95 \pm 0.01$ & $1.96 \pm 0.01$ & $358.8 \pm 0.1$ & \\
1998.5 & 1007.6 & HST/Nicmos    &HK(d)&$-0.038 \pm 0.001$&$1.960 \pm 0.006$ & $1.960 \pm 0.006$&$ 358.89  \pm 0.03$ & Wei00 \\
\multicolumn{4}{r}{corrected:} & $-0.038 \pm 0.001$&$1.960 \pm 0.006$ & $1.960 \pm 0.006$&$ 358.89  \pm 0.03$ & \\
2000.1 & 1589.6 & Keck/KCam     & H   & n/a (g) & n/a (g) & $1.956  \pm 0.012  $& n/a (g)                & Mac01 \\
\multicolumn{4}{r}{corrected:} & n/a (g) & n/a(g) & $1.931 \pm 0.012$ & n/a (g) &  \\
2000.1 & 1596.6 & VLT/FORS2     & I   & $-0.091 \pm 0.037$ & $1.950 \pm 0.034$ & $1.952  \pm 0.05   $&$ 357.3   \pm 1.2$  & Neu00 \\
\multicolumn{4}{r}{corrected:} & $-0.102 \pm 0.037$ & $1.925 \pm 0.034$ & $1.928 \pm 0.04$ & $357.0 \pm 1.3$ & \\
2000.1 & 1597.6 & Keck/KCam     & JHK & $-0.0286 \pm 0.0467$ & $1.9538 \pm 0.0320$ & $1.954  \pm 0.008  $&$359.16   \pm 1.08$ & Bra03 \\
\multicolumn{4}{r}{corrected:} & $-0.0386 \pm 0.0467$ & $1.9288 \pm 0.0321$ & $1.9292 \pm 0.057$ & $358.85 \pm 1.18$ & \\
2000.3 & 1651.6 & VLT/ISAAC     & (d) &$-0.093 \pm 0.097$&$1.967 \pm 0.096 $ & $1.969 \pm 0.091 $&$ 356.8 \pm 3.0$ & Neu00 \\
\multicolumn{4}{r}{corrected:} & $-0.103 \pm 0.097$ & $1.944 \pm 0.096$ & $1.947 \pm 0.091$ & $357.0 \pm 3.1$ & \\
2001.2 & 1965.0 & Gem/Hokupaa   & H (e)  & n/a (g) & n/a (g) & $1.9357 \pm 0.0107 $& n/a (g)                & Neu01 \\
\multicolumn{4}{r}{corrected:} & n/a (g) & n/a (g) & $1.9157 \pm 0.0109$ & n/a (g) &  \\
2003.1 & 2687.8 & VLT/NACO      & K$_{s}$ & $-0.076 \pm 0.013$&$1.921 \pm 0.020$&$1.923 \pm 0.020$&$ 357.74 \pm 0.38$ & Mas05 (h) \\
\multicolumn{4}{r}{corrected:} & $-0.066 \pm 0.013$ & $1.924 \pm 0.020$ & $1.925 \pm 0.020$ & $358.04 \pm 0.48$ & \\
2007.5 & 4291.0 & VLT/NACO      & K$_{s}$ & $-0.123 \pm 0.013$&$1.897 \pm 0.011$&$1.901 \pm 0.011$&$ 356.28 \pm 0.40$ & this work \\
\multicolumn{4}{r}{corrected:} & $-0.123 \pm 0.013$ & $1.882 \pm 0.011$ & $1.886 \pm 0.011$ & $356.26 \pm 0.58$ & \\
2008.4 & 4630.5 & VLT/NACO      & K$_{s}$ & $-0.130 \pm 0.013$&$1.887 \pm 0.012$&$1.892 \pm 0.012$&$ 356.07 \pm 0.40$ & this work \\  
\multicolumn{4}{r}{corrected:} & $-0.125 \pm 0.013$ & $1.883 \pm 0.012$ & $1.887 \pm 0.012$ & $356.20 \pm 0.50$ & \\ \hline
\end{tabular}

Remarks: (a) {\em Corrected} values are separations and PA between A and B after correction
for photocenter shift of Aa+b due to orbit of Ab around Aa (K07).
Separation in $\alpha$ is negative for B west of A,
and separation in $\delta$ is positive for B north of A.
(b) PA is position angle measured from north over east to south.
(c) References are Lowrance et al. 1999 (Low99), 
Macintosh et al. 2001 (Mac01), Weintraub et al. 2000 (Wei00),
Neuh\"auser et al. 2000 (Neu00), Webb et al. 1999 (Webb99), 
Neuh\"auser et al. 2001 (Neu01), 
Brandeker et al. 2003 (Bra03), and Masciadri et al. 2005 (Mas05).
(d) Narrow band filter(s). 
(e) With Wollaston polarimeter.
(f) Values (listed in the first row for Low99) are those given
by Wei00 according to priv. comm. with P. Lowrance,
corrected compared to those published in Low99.
(g) Values not available (not published).
(h) Mas05 list neither separation nor PA, data reduced by us.

\end{table*}

Figures 1 to 3 show how separation and PA change with time.
We always plot the {\em corrected} values for separations and PA.
The data are inconsistent with B being a non-moving background object
by many $\sigma$, but fully consistent with TWA 5 A+B being a 
common-proper motion pair, as known before
(Neuh\"auser et al. 2000, Brandeker et al. 2003). 
We show in Figs. 1 and 2 the maximum possible orbital motion for a circular
orbit of TWA 5 B around A.
We use $0.71 \pm 0.14$~M$_{\odot}$ as total mass towards TWA 5 A (K07).
The observed change in PA (Fig. 1) is close to the maximum expected for a circular orbit,
which would indicate that either the orbital plane is seen nearly pole-on and/or that
the orbit is eccentric. In the former case, there should be almost no change in separation;
however, the separation also changes significantly, so that we can conclude that
the orbit of B around A is eccentric (and maybe also inclined).

The distance towards TWA 5 is not measured as parallax, 
but Hipparcos has measured the parallax of three to five members of
the TW Hya association (TWA), where TWA 5 is a certain member,
namely for TWA 1, 4, 9, 11, and 19.
The use of the values for TWA 9 and 19 is dubious due to binarity and
uncertain membership, respectively (Mamajek 2005). The weighted mean of the 
distances towards TWA 1, 4, and 11 is $61.6 \pm 2.8$ pc.
There were also a few measurements of the distance towards TWA 27 (2M1207),
namely $58 \pm 7.0$ pc (Biller \& Close 2007),
$54.0 \pm 3.0$ pc (Gizis et al. 2007), and
$52.4 \pm 1.1$ pc (Ducourant et al. 2008), along with an indirect 
determination through its proper motion to be $59 \pm 7$ pc (Song et al. 2006), 
The weighted mean of the three ground-based distances 
of TWA 27 (2M1207) is then $52.7 \pm 1.0$ pc.
The weighted mean of the distances towards TWA 1, 4, 11, and 27
is then $53.73 \pm 0.94$ pc.
The ground-based trigonometric distance towards TWA 22 is only $17.5 \pm 0.2$ pc
(Teixeira et al. 2009), so that it might be dubious as to whether TWA 22 is a member.
The mean distance when including one to three of the stars TWA 9, 19, and 22
(in addition to TWA 1, 4, 11, and 27) lies then between 48 and 65 pc.
The weighted mean would be much smaller, when including TWA 22 owing to
its very smaller error ($\pm 0.2$ pc). 
Mamajek (2005) reports a kinematical distance of $44 \pm 4$ pc 
for TWA 5, but when using the uncertain TWA 9 parallax estimate in his 2D convergent point method. 
Torres et al. (2008) report a distance of 45 pc using a 3D kinematical convergent point method.
Given the distance range of TWA members with trigonometric distance determination 
between 17 and 104 pc, given the mean distance of the best values of around $\sim 50$ pc,
and also given the wide spread of TWA members on the sky,
the individual distance towards TWA 5 of $44 \pm 4$ pc (and $\sim 45$ pc) seems plausible,
so it is used here (as in K07).

The most precise measurement of the separation between A and B is
$1.960 \pm 0.006 ^{\prime \prime}$ (epoch 1998.5).
With a distance of $44 \pm 4$ pc,
the projected physical separation between TWA 5 B and Aa+b is $86.2 \pm 4.0$ AU,
which would be the semi-major axis for a circular orbit.
For a total system mass of $0.71 \pm 0.14$~M$_{\odot}$ (K07), 
the orbital period would then be $\sim 950$ yrs (for a circular orbit).
For our best fit (geometric) orbit as an ellipse (with e=0.45),
and given the current location of B on its orbit around A,
the semi-major axis is $\sim 100$ AU (at 44 pc) and the orbital
period then $\sim 1200$ yrs.

\section{Interpretation}

The separation between TWA 5 A and B in both $\alpha$ and $\delta$
change {\em slightly} with time. This is typical of a common-proper motion
pair, where orbital motion is seen. Because of orbital motion, $\alpha$ and/or
$\delta$ do not stay exactly constant, but can change slightly.
However, a constant change in the separation in both $\alpha$ and $\delta$
would also be consistent with B being a moving background object:
If B moves slightly, but with a proper motion different than A,
then we would see a slight change in separation.
If both objects are orbiting each other, we should see curvature 
in the orbital motion after some time. Such a curvature could 
in principle also be consistent with hyperbolic motion.
All these alternatives are also possible in all the other substellar companions detected
directly so far.
Curvature in orbital motion has been shown for none of them, yet.
Even if youth indicators are detected in the spectrum of an apparently
co-moving substellar companion (like low gravity or accretion), the object
could still be an independent member of the same young association as the
primary object. In such a case, the expected difference in proper motion
could reach a few mas/yr, the typical velocity dispersion in young clusters
(see e.g. Herbig \& Jones 1979, Mugrauer \& Neuh\"auser 2005).
A final proof of companionship could be curvature in orbital motion,
consistent with a circular or elliptic orbit.

Given the few data points with large error bars, 
we cannot yet fit a full physical orbit for the motion of B around A. 
However, we can try to fit a geometric orbit by testing different hypotheses 
by estimating the probabilities for different models
of motion of B relative to A with Bayesian statistics:
a constant change in separation in $\alpha$ and $\delta$
as if B were a background object with proper motion
different from A,
a circular orbit of B around A,
an elliptic orbit of B around A,
hyperbolic motion of B wrt A (i.e. B being ejected),
and A and B as an exactly co-moving pair; i.e., both objects
having the same motion in $\alpha$ and $\delta$, so that the
separation in both $\alpha$ and $\delta$ remain constant.
Assuming that a priori those five models are equally probable and
provide a complete set of hypotheses, we calculated global likelihoods
(Bayesian model comparison, see Gregory 2005). 

According to these calculations, an elliptic orbit is most likely,
with a probability for this hypothesis being 0.63, which is not
yet very significant, but 2 to 3 times larger than the probabilities for the 
two other likely hypotheses (constant change and hyperbola),
and even many times larger
than for circular orbit and exactly co-moving.
The best fit ellipse has a position angle of $106^{\circ}$ and
an eccentricity of $e \simeq 0.45$.

\begin{figure}
\includegraphics[angle=0,width=1\hsize]{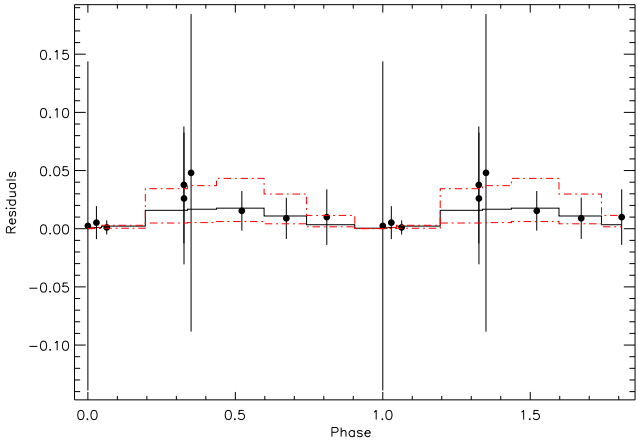}
\caption{Phase-folded residuals (in arc sec) from the (best) 
geometric fit (elliptic orbit of TWA 5 B around the photocenter 
of Aa+b, similar to Fig. 3, but for {\em un}corrected data) with 
data points with error bars ({\em un}corrected data from Table 1 
after subtraction from best-fit orbit), 
best-fitting sinusoid for residuals (full black line),
and its $\pm 1 \sigma$ limits (as dashed-dotted red lines).
Period ($5.72 \pm 1.14$ yr), amplitude ($18 \pm 9$ mas), and phase zero point 
(epoch 1998, at phase 0.0 and 1.0) of these residuals are fully
consistent with the orbit of Ab around Aa found in K07,
see Sect. 4.}
\end{figure}

Hence, we have weak evidence that an elliptic orbit is 
more probable than any other model, hence evidence for curvature
in the orbital motion of B around A. If this can be confirmed,
TWA 5 A+B would be the first substellar companion outside the solar 
system, where such evidence is reported from direct imaging observations.

Then, we can also try to investigate whether we can detect a small periodic
wobble in the separation of B from A, which would stem from the expected 
photocenter shift of the close pair Aa+b. This wobble should be seen in
periodic residuals to the best fit (ellipse of B around A) in the {\em un}corrected
data from Table 1. The period would then give the orbital period of Aa+b,
and the amplitude would yield the total mass of Aa+b. For the uncorrected 
(i.e. directly observed) data in Table 1, we also obtain an ellipse
as the best geometric fit to the data by Bayesian statistics. As can be seen
in Fig. 4, the residuals to that best fit do show a small-amplitude
(periodic) sinusoid. We searched for the periodicity only in a small
window of 3-8 years (around the known orbital period of 5.94 yr)
and detected a best-fit period of $\sim 5.72 \pm 1.14$ yr. The (half-)amplitude 
of the sinusoidal wobble is $18 \pm 9$ mas.
Both values are close to the values from the orbital fit of Aa+b
in K07, where they give $5.94 \pm 0.09$ yr period. 
Given the semi-major axis, eccentricity, and inclination,
the maximum photocenter shift on the sky (for equal
brightness of Aa and Ab) is $\sim 28$ mas (their Fig. 3 and Table 2).
Our residuals are close to zero (minimum of residuals) at epoch $\sim 1998$,
roughly six years before the epoch of closest approach given
in K07. Periastron and closest separation
on sky are very close together (K07).
As a result, our data are consistent with the orbit found in K07.

Even though our data have lower angular resolution (8m VLT)
compared to K07 using the 10m Keck, so that the inner
TWA 5 Aa+b pair is unresolved in our NACO data,
we can detect and measure the photocenter wobble of TWA 5 Aa+b in the
separation changes between A and B. In principle, we could also
measure the total mass of the TWA 5 Aa+b pair; however, we refrain from doing
so, because our images do no resolve for the close pair and, thus, have much lower 
precision compared to K07.
This method should now also be applicable to other cases,
at least for detecting close pairs, possibly 
also for testing and calibrating pre-main sequence models.


\begin{acknowledgements}
We would like to thank the ESO Paranal Team and ESO Users Support group.
We are grateful to Markus Mugrauer for useful discussion.
TOBS would like to thank Evangelisches Studienwerk e.V. Villigst for financial support.
NV acknowledges support by FONDECYT grant 1061199.
RN wishes to acknowledge general support from the German National Science Foundation
(Deutsche Forschungsgemeinschaft, DFG) grants NE 515/13-1 and 13-2 (until 2006),
as well as NE 515/23-1, 30-1, and 32-1 (since 2007).
VVH would like to acknowledge support from the
DFG in the Sonderforschungsbereich SFB TR 7 on Gravitatinal Wave Astronomy.
\end{acknowledgements}


\end{document}